\documentclass[11pt]{article}
\usepackage{moriond,epsfig}

\def\Journal#1#2#3#4{{#1} {\bf #2}, #3 (#4)}

\def\PRD{{\em Phys. Rev.} D}

\def\CQG{\em Class. Quant. Grav.}

\def\be{\begin{equation}}
\def\ee{\end{equation}}
\def\bea{\begin{eqnarray}}
\def\eea{\end{eqnarray}}

\begin{document}
\vspace*{4cm}
\title{Searching for Gravitational-Wave Bursts with LIGO}

\author{ K. A. Thorne, for the LIGO Scientific Collaboration (LSC)}

\address{Center for Gravitational Wave Physics \\ Institute for  
Gravitational Physics and Geometry \\ Pennsylvania State University\\
University Park, PA 16802 United States}

\maketitle\abstracts{
We present recent results from searches by the LIGO Science
Collaboration for bursts of gravitational-wave radiation, as well as
the status of other ongoing searches.
These include directed searches for bursts associated with observed  
sources (gamma-ray bursts, soft gamma repeaters) and untriggered  
searches for bursts from unknown sources. We also present the status  
of some newer investigations, such as coherent network methods.
We show methods for interpreting our search results in terms of  
astrophysical source distributions that improve their accessibility  
to the wider community.
}

\section{Introduction}

The Laser Interferometer Gravitational Wave Observatory (LIGO) is in
the middle of a lengthy science run (S5) in the search for
gravitational-wave (GW) signals.  One class of signals are
short-duration ($< 1$ sec) ``bursts'' of gravitational-wave
energy. The LIGO Science Collaboration (LSC), an international
organization of researchers working with the LIGO~\cite{ligo} and
GEO~600~\cite{geo} detectors, is continuing searches for these GW
bursts started in previous science runs.  Section~\ref{sec:recent}  
reviews
recent progress and results from LIGO-only searches.  The remainder
of the paper discusses some examples of new analysis directions being
pursued by members of the LSC.  Section~\ref{sec:coherent}
covers new work on network-based burst searches, looking towards the
addition of data from GEO 600, Virgo~\cite{virgo}, and eventually other
observatories.  The last section covers methods for presenting GW  
burst search results in
terms of rate limits for models of astrophysical source distributions.

\section{Recent LIGO GW Burst Searches} \label{sec:recent}
Unlike the well-modeled waveforms for GW signals from pulsars and the  
inspiral phase of binary compact object mergers, GW bursts are poorly  
modeled at present.  Searches for
GW burst signals thus must remain sensitive to a large range of  
signal waveforms.  We divide
the searches into two classes. One class are untriggered searches that
examine all sky locations at all observation times. The other class  
are directed searches for GW burst signals associated with  
astronomically-identified source candidates such as Gamma-Ray Bursts  
(GRBs) of known sky location and observation time.

\subsection{All-Sky Untriggered Burst Search}
The initial untriggered burst search for LIGO run S5 uses the same  
approach as was used for runs S2, S3, and S4~\cite{bursts4}. The  
search starts with a wavelet decomposition of the gravitational-wave  
channel data from each detector separately into time-frequency maps.  
Samples (``pixels'') from these
maps that have excess signal power relative to the background are  
identified.  Such pixel clusters that are coincident in time and  
frequency
between all three LIGO interferometers are selected as candidate  
triggers for further analysis.  The candidate triggers must then
pass a set of signal consistency tests.  Based around pair-wise cross- 
correlation tests, these confirm that consistent waveforms
and amplitudes are seen in all interferometers.  These same methods  
are used to measure background rates by processing data
with many artificial time shifts between the two LIGO sites in Hanford
(Washington) and Livingston (Louisiana).

The LIGO-only burst GW analysis can have significant backgrounds from  
non-Gaussian transients.  A particular problem are environmental  
transients at the Hanford site. These can induce simultaneous large- 
amplitude signals in the co-located interferometers (labeled H1 and  
H2) at that location. Detailed studies of Data Quality (DQ) are  
required to identify and define time intervals when such problems are  
present. This work is assisted by the large number of auxiliary  
channels of interferometer and environmental sensor data that are  
recorded during science operation.  Longer-duration time periods that  
have known artifacts or unreliable interferometer data are flagged as  
DQ Period Vetoes. Short-duration transient events in the auxiliary  
channels that are found to be associated with events in the GW  
channels are flagged as Auxiliary-Channel Event Vetoes.  Both veto  
classes are used to reject GW Burst triggers that coincide with  
them.  These vetoes help remove any large-amplitude outliers in the  
final GW Burst trigger samples~\cite{bursts4}.

A detailed analysis of untriggered burst search results from the  
early part of S5 operation is being completed for publication. We  
note that our searches in the previous LIGO runs (S1~\cite{bursts1},  
S2~\cite{bursts2}, S3~\cite{bursts3} and S4~\cite{bursts4}) did not  
see any GW burst signals. The S5 run has both greater sensitivity  
than previous runs and at least 10 times the observation time. As in  
S4, this initial S5 all-sky GW burst search is tuned by bursts $\ll 1 
$ sec in duration over a frequency range of 64-1600 Hz.
As there are few well-modeled waveforms for bursts from theoretical  
studies, we use `ad-hoc' waveforms such as Gaussian-envelope sine- 
waves (sine-Gaussians) and Gaussians that mimic the expected  
transient response to such bursts. We measure our sensitivity to such  
ad-hoc waveforms in terms of their root-sum-squared amplitude ($h_ 
{rss}$) which
is in units of $\mbox{strain}/\sqrt{\mbox{Hz}}$ defined as
\begin{equation}
h_{rss} = \sqrt{\int(|h_{+}(t)|^{2} + |h_{\times}(t)|^{2})dt} \, .
\label{eq:hrss}
\end{equation}

\begin{figure}
\begin{center}
\epsfig{figure=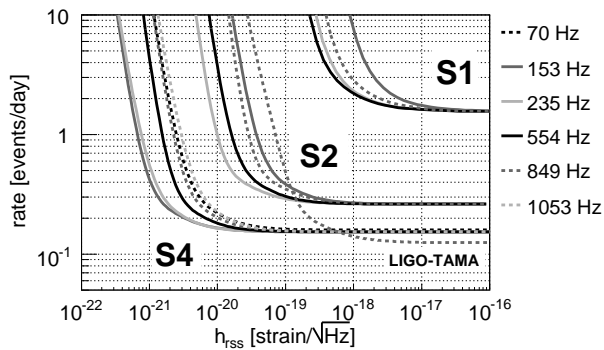, width=3.75 in}
\caption{Exclusion diagram (rate limit at 90\% confidence level, as a  
function of signal amplitude) for sine-Gaussian simulated
waveforms for S1, S2 and S4 LIGO GW burst analyses (No rate limit quoted in S3 analysis).
\label{fig:excSG}}
\end{center}
\end{figure}

The ``efficiency'' of an analysis is the probability that it will
successfully identify a signal with certain specified parameters.
For an all-sky search, we use a Monte Carlo approach to evaluate the
efficiency for each of these waveforms as a function of amplitude,
averaging over sky position and polarization angle.
This information is used to derive exclusion diagrams that place  
bounds on the event rate as a function of $h_{rss}$.
This is shown in Fig.~\ref{fig:excSG}, taken from our recent S4 paper~ 
\cite{bursts4}.
In our early S5 analysis, we are achieving detection sensitivities of  
$h_{rss} < 10^{-21} \, \mbox{strain}/\sqrt{\mbox{Hz}}$ for some of  
the ad-hoc waveforms considered.
These instrumental sensitivities can also be converted to
corresponding energy emission sensitivity~\cite{kr}. Assuming (for
simplicity) isotropic emission at a distance $R$ by GW bursts
% with $h_{+}$ only polarization
with sine-Gaussian waveforms, we have
\begin{equation}
E_{GW} = (2.1 M_{\odot}c^{2}) \left(\frac{R}{100 \, \mbox{Mpc}}\right) 
^{2} \left(\frac{f}{100\,\mbox{Hz}}\right)^{2}
\left(\frac{h_{rss}}{10^{-21} \,\mbox{Hz}^{-1/2}}\right)^{2}
\end{equation}
During the early part of S5, we are sensitive to $E_{GW} \sim 0.1 M_ 
{\odot}c^{2}$ at a distance of 20 Mpc for $f = 153$ Hz sine-Gaussians.

\subsection{GRB-triggered Burst Search}
We have completed a search for short-duration GW bursts that are  
coincident with Gamma-Ray Bursts (GRBs) from the data in several  
previous LIGO science runs (S2, S3 and S4).  This analysis used pair-wise 
cross-correlation of signals from two LIGO interferometers, as  
used in our search for gravitational waves associated with 
GRB030209~\cite{grb030329}.  This approach increased the observation time over  
that available when all three LIGO interferometers were in science  
mode.  The search targeted bursts with durations 1 to 100 ms over a  
bandwidth of 40--2000 Hz.
%
%  - Omitted due to lack of published result --
%  There were no GW burst signals found that were associated with  
%          the 39 GRBs during the S2, S3 and S4 runs.
%  
The sensitivity of this  GRB search is similar to that of the untriggered search.
As well as setting limits on GW bursts from individual GRBs, it was  
demonstrated that multiple GRBs can be combined to
obtain an upper limit on GRB population parameters.
During S5, there have been about 10 GRBs per month.  Thus, the GRB  
sample that will be used in the S5 analysis will be much larger.

\subsection{SGR 1806--20 Search}
We have also completed a search for GW signals associated with the  
Soft Gamma-Ray Repeater (SGR) 1806--20.  This SGR had a record  
flare on December 27, 2004~\cite{sgr1806}.  During this flare, quasi-periodic 
oscillations (QPOs) were seen in X-ray data from the RHESSI  
and RXTE satellites.  These QPOs lasted for hundreds of seconds.   
During this flare, only one LIGO detector (H1) was observing.  A band-limited 
excess-power search was conducted for quasi-periodic GW  
signals coincident with the flare~\cite{sgrsearch}.  No evidence was  
found for GW signals associated with the QPOs.  Our sensitivity to  
the 92.5 Hz QPO was $E_{GW} \sim 10^{-7} \mbox{ to } 10^{-8} M_{\odot}$, 
based on the 5--10 kpc distance of SGR 1806--20.  This is  
comparable to the total electro-magnetic energy emitted in the flare.

\section{Coherent GW Burst Searches} \label{sec:coherent}
The existing LIGO all-sky untriggered and GRB triggered burst search  
pipelines have been operating continuously on the acquired science- 
mode data since the start of the S5 run.  These provide for the  
chance of prompt detection of GW bursts, enabling timely follow-up  
and investigation.  The results are also used to provide  
identification of false signals from transients, speeding up the data  
quality and auxiliary-channel veto studies.

In searching for GW bursts, the community is adopting an approach  
that might be termed ``The Network is the Observatory''.   The  
benefit of having observatories at multiple, widely-separated  
locations cannot be stressed enough. While LIGO-only burst searches  
have been fruitful, they require intense investigations for  
environmental and interferometer transients to remove backgrounds.   
Our previous LSC analyses have not made full use of the constraints  
that a network of sites can jointly make simultaneously on $h_{+}$  
and $h_{\times}$ waveforms. Specifically, the GW burst searches must  
prepare for the inclusion of data from the GEO 600 and Virgo  
observatories, and others in the future.

We fully expect to move from the era of upper limits to that of  
detection.  In moving to detection, GW burst searches need to extract  
the waveform of the signals that are detected.  Such waveforms can be  
compared to those from theoretical predictions for potential  
identification of the source type.  We may also provide interpretations of our 
GW burst results in terms of rates from astrophysical source distributions.

\subsection{Coherent Network Burst Searches}
A coherent method for GW burst searches was first proposed by G\"{u}rsel 
and Tinto~\cite{GuerTint}, where they combined the
detector responses into a functional.  This functional minimizes in  
the direction of the source and allows extraction of
both the source coordinates and the two polarization components of  
the burst signal waveform.  Flanagan and Hughes~\cite{FlanHugh}
expanded this to the maximization of a likelihood functional over the  
space of all waveforms.  Using simulated data, Arnaud {\it et al}~ 
\cite{Arnaud} found that coherent methods were more efficient than coincidence methods  
for burst signals, in an exploration of their statistical performance.
Within the LSC, there has been substantial work to develop searches  
that evaluate the network's composite response to GW burst signal.
Such ``coherent network'' techniques can accommodate arbitrary  
networks of detectors and  will be among the methods used by the 
LSC for the analysis of the S5 data.

To describe coherent network burst searches, we will follow the  
presentation in Klimenko {\it et al}~\cite{Klimenko}.
In the coordinate frame associated with the wave (termed the wave  
frame), a gravitational wave propagates in the direction of the $z$ axis.
For a specific source, the $z$-axis is defined by the source's  
location on the sky in terms of angles $\theta$ and $\phi$.  The wave  
can be
described with the $h_{+}$ and $h_{\times}$ waveforms representing  
the two independent polarization components of the wave.
In describing the network analysis, we will use complex waveforms  
defined as
\begin{eqnarray}
u(t) & = & h_{+}(t) + ih_{\times}(t) \\
\tilde{u}(t) & = & h_{+}(t) - ih_{\times}(t)
\end{eqnarray}
We use a tilde($\tilde{\,}$) to indicate the complex conjugate. These complex  
waveforms are eigenstates of the
rotations about the $z$ axis in the wave frame.

The response $\xi(u)$ of an individual detector can be expressed conveniently 
in terms of these complex waveforms:
\begin{equation}
\xi(u) = \tilde{A} u + A \tilde{u}
\end{equation}
where $A$ and $\tilde{A}$ are complex expressions of the standard  
antenna patterns. We note that
this detector response is invariant under rotations $R_{z}$ in the  
wave frame.
This can be extended to a network of detectors, where the response  
from each detector is weighted
by its noise variance $\sigma^{2}$.  We combine the per-detector antenna patterns into 
the {\it network  antenna patterns} 
\begin{equation}
g_{r} = \sum_{k=1}^{K}\frac{A_{k}\tilde{A}_{k}}{\sigma_{k}^{2}}, \;
g_{c} = \sum_{k=1}^{K} \frac{A_{k}^{2}}{\sigma_{k}^{2}} 
\end{equation}
where $g_{r}$ is real and $g_{c}$ is complex. The analogous {\it  network response} $R(u)$ 
is expressed in terms of these patterns:
\begin{equation}
R(u) = g_{r} u + g_{c} \tilde{u}
\label{eq:Ru}
\end{equation}
There is also the network output time series $X$ that combines the  
output time-series $x_{k}$ from each detector
\begin{equation}
X = \sum_{k=1}^{K} \frac{x_{k}A_{k}}{\sigma_{k}^{2}}
\end{equation}

The equations for the GW waveforms from the network are obtained by variation of the  
likelihood functional.  This results in two linear equations for $u$ and $\tilde{u}$
\begin{eqnarray}
X & = & g_{r} u + g_{c} \tilde{u} \\
\tilde{X} & = & g_{r} \tilde{u} + \tilde{g}_{c} u
\end{eqnarray}
These can be written in matrix form as
\begin{equation}
\left[\begin{array}{ c }
	Re(X) \\ Im(X)
\end{array} \right] = M_{R}
\left[\begin{array}{ c }
	h_{+} \\ h_{\times}
\end{array} \right]
\end{equation}
where $M_{R}$ is the network response matrix.

The invariance of the response to an arbitrary rotation $R_{z}(\psi)$  
through an angle $\psi$ allows us to
select the rotation in which both network antenna patterns $g_{r}$  
and $g_{c}$ are real and positively defined.
This simplifies the network response to
\begin{equation}
R = (g_{r} + |g_{c}|) h_{1} + i(g_{r} - |g_{c}|) h_{2}
\end{equation}
where $h_{1}$ and $h_{2}$ are the real and imaginary components of  
the signal.  That leads to a diagonal form for the network response matrix:
\begin{equation}
M_{R} = g
\left( \begin{array}{ c c }
	1 & 0 \\ 0 & \epsilon
\end{array} \right)
\end{equation}
The coefficient $g$ characterizes the network sensitivity to the $h_{1}$ wave.  The sensitivity to the
$h_{2}$ wave is $\epsilon g$, where $\epsilon$ is the network  
alignment factor.  For most sources,
the $h_{1}$ and $h_{2}$ components should have similar amplitudes.

However, there is a problem in the use of coherent network approaches  
that is most acute for a network of two detector locations, such as
the LIGO-only configuration with detectors at Hanford and  
Livingston.  It has been shown~\cite{Klimenko} that if the detectors  
are even slightly misaligned, the normal likelihood statistic becomes  
insensitive to the
cross-correlation between the detectors.  This results in the network alignment factor $\epsilon$ being 
$\ll 1$ for most
sky location angles $\theta$ and $\phi$, and hence the network being  
insensitive to the $h_{2}$ wave component.
This problem lessens somewhat as more detectors, such as GEO 600 and  
Virgo, are added, but does not disappear.

\subsection{Application to Untriggered Burst Search}
One method that has been developed to deal with the relative  
insensitivity to the $h_2$ component is the ``constraint likelihood''~\cite{Klimenko}.
This applies a ``soft constraint'' on the solutions that penalizes  
the unphysical solutions with $h_{1} = 0$ that would be
consistent with those produced by noise.   This sacrifices a small  
fraction of the GW signals but enhances efficiency for the
rest of the sources.

The existing wavelet-based search has been converted into an all-sky  
coherent network search using
this ``constraint likelihood''  technique.  It can handle arbitrary  
networks of detectors and has been tested on
existing data from the LIGO and GEO 600 detectors.   This all-sky  
coherent network search also divides the detected energy from all  
detectors into
coherent and incoherent components.  A cut on the network correlation  
(coherent / total) further
removes backgrounds from single-detector transients.  When compared  
to our existing
all-sky search method for S5, this coherent network search achieves  
equal or better
sensitivity with a very low background rate.
Other coherent search methods have also been explored and/or
implemented~\cite{ExcessPower,WenSchutz,ChatterjiEtAl}.

\subsection{Application to Triggered Burst Search}
Additional methods for handling the $h_1$ and $h_2$ polarization
components in the likelihood have been
studied~\cite{Mohanty,Rakhmanov}.
It was noted that problems arise when the inversion of the detector  
response to obtain the waveforms is {\it ill-posed} due to ``rank deficiency''.   
This can be solved using many types of regularization.

The method of Tikhonov regularization~\cite{Rakhmanov} is used in a  
new triggered coherent network analysis~\cite{Hayama} developed
by the LSC for S5 and subsequent data sets.  Because there is prior  
knowledge of the sky locations, and fewer sources
than in the untriggered analysis, more computationally intensive  
methods can be used. In fact, some additional
network analysis methods are under development for the triggered  
burst search.

%Coherent network searches have the inherent ability to extract the  
% waveform of the detected signal.
%These are most successful when either the signal has a large SNR,  
% or testing for the
% presence of waveforms of specific morphologies.

\section{Astrophysical Interpretation}

The existing GW burst search results from LIGO (See
Fig.~\ref{fig:excSG}) have been reported in terms of detector-centric
``Rate vs. Strength'' exclusion curves.  These methods say nothing
about the sources of GW bursts or about the absolute rate of source
events.  The ``Rate'', typically in events/day, is only meaningful if
all events are assumed to have the same ``Strength'', {\it i.e.}\ GW
strain amplitude at the Earth, expressed in terms of $h_{rss}$.
%% in units of $ \mbox{strain}/\sqrt{\mbox{Hz}}$.
This strength parameter reveals little about
source quantities such as the absolute luminosity in terms of emitted
gravitational-wave energy.

Alternatively, we could report results in terms of rates from a source
population as a function of the intrinsic energy radiated.  We note
that interpretation, astrophysical or otherwise, is always in terms of
a model.  The components of such a model would be the source
population distribution and the source strain energy spectrum
appropriate for GW burst searches.  We also need to add in the
observation schedule. This schedule is the sidereal time associated with the
data that is analyzed, that provides the detector pointing
relative to the source population distribution.

We wish to report results in terms of their {\em Astrophysical} Rate
vs. Strength. The Astrophysical Rate is the event rate in the source
population.  The Astrophysical Strength is an astrophysically
meaningful amplitude parameter such as the radiated energy.  We
express the bound on the astrophysical rate vs strength
\begin{equation}
R(E) = \frac{k}{T_{obs}\epsilon(E)}
\label{eq:ratestr}
\end{equation}
where the constant $k$ is set by the number of observed events (2.3  
for no observed events), $T_{obs}$ is the total observation time and 
$\epsilon(E)$
is the efficiency in the population.  The efficiency in the  
population is the ratio of the expected number of observed sources over the total number of  
sources.  The expected number of observed sources is the integral of the source rate distribution over  
detection efficiency and observation schedule.  The total number of sources is the integral of  
the source rate distribution over the observation schedule alone.    
The source rate distribution will be a function of location,  orientation and luminosity.

\subsection{Example of Astrophysical Interpretation}
It is best to illustrate what an astrophysical interpretation means  
with an example.  We start by choosing a source population.  We will  
assume the source population traces out the old stellar population.   
We will thus use a Milky Way galactic model with a thin disk, thick  
disk, bulge, bar and halo that are characteristic of the observed  
white dwarf population.  For a source model, we will assume an  
implusive event that involves stellar-mass compact objects.  These  
events could be supernovae, accretion-induced collapses (AIC), etc.  
We will assume axisymmetric GW bursts and ``standard candle''  
amplitudes, {\it i.e.}\ each source has the same absolute luminosity  
in GW energy.  For the network of detectors, we assume  
interferometers at the
LIGO Hanford, LIGO Livingston and Virgo Cascina sites. Each site has  
an interferometer (labeled H1, L1 and V1)
with a detection sensitivity characterized as a step-function at 
$h_{rss} \sim 10^{-20} \,\mbox{Hz}^{-1/2}$. The LIGO Hanford site has an  
additional interferometer (H2) that has 
$h_{rss} \sim 2 \times 10^{-20} \,\mbox{Hz}^{-1/2}$ due to being half as long as H1.
For this example, we will assume a 100\% observation schedule,  
implying uniform coverage in sidereal time.

\begin{figure}
\begin{center}
\epsfig{figure=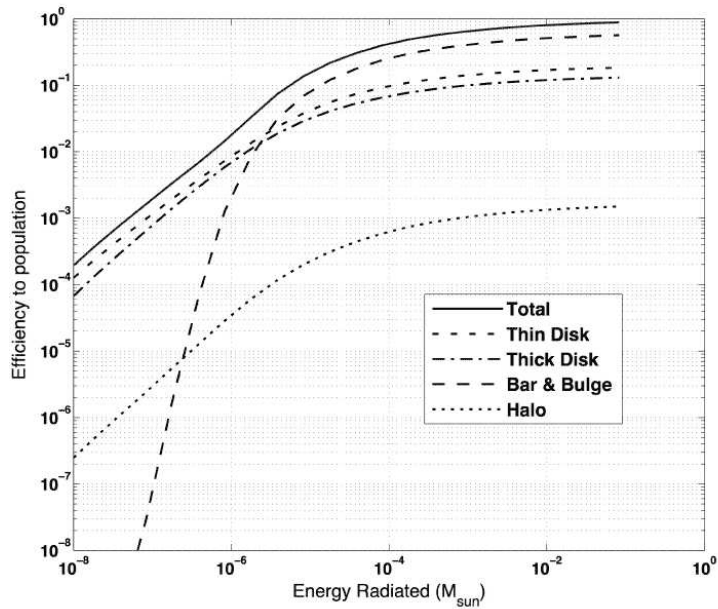, width=4.25 in}
\caption{Detection efficiency for the example galactic source
population model described in the text. Note
how disk components dominate at low levels of intrinsic radiated  
energy, while
bar and bulge components dominate for larger levels of radiated energy.
\label{fig:effpop}}
\end{center}
\end{figure}

First we calculate the efficiency to the population $\epsilon(E)$ as  
a function of the energy radiated ($E$).
This is shown in Fig.~\ref{fig:effpop} which has the efficiency  
broken down by contributions from each galactic model component to  
the total as a function of radiated energy in solar masses.
Note in our example that for radiated energy above $10^{-5}$ solar masses,  
contributions from the bar and bulge components dominate, while below  
$10^{-6}$ solar masses, the thin and thick disk components dominate.

\begin{figure}
\begin{center}
\epsfig{figure=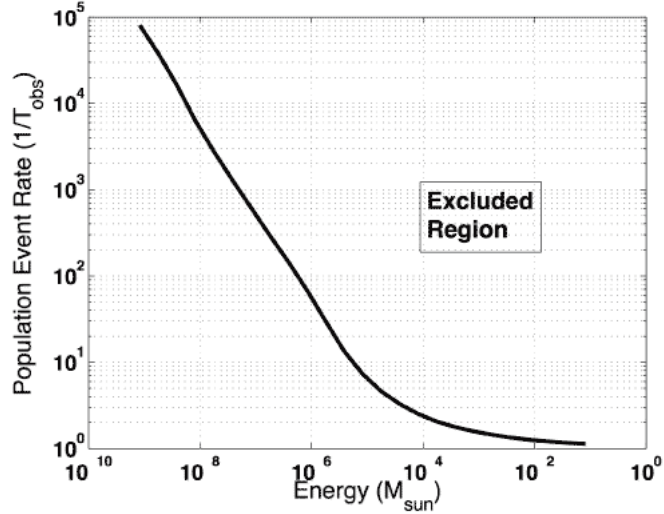,width = 3.75 in}
\caption{Bound on Population Event Rate as a function of Radiated  
Energy.
\label{fig:ratestr}}
\end{center}
\end{figure}

The efficiency to the population is used to derive the bound on  
population rate vs. strength. This is shown in Fig.~\ref{fig:ratestr}  
for the example.
As different components dominate, the shape of the exclusion curve  
changes.

\section{Conclusions}
The LIGO-based burst searches are well established and already  
processing the
data from the current S5 science run.  New network-based techniques  
have been
developed that provide enhanced detection sensitivity and background  
rejection.
These methods show our preparation for joint observation with the  
Virgo observatory.
The introduction of results interpretation in terms of astrophysical  
source distributions
improves their accessibility to the astronomy and astrophysics  
communities.

\section*{Acknowledgments}
This work was supported by the National Science Foundation under  
grant 570001592,A03.
In addition, the LIGO Scientific Collaboration
gratefully acknowledges the support of the United States
National Science Foundation for the construction and operation of
the LIGO Laboratory and the Particle Physics and Astronomy Research
Council of the United Kingdom, the Max-Planck-Society and the State
of Niedersachsen/Germany for support of the construction and
operation of the GEO 600 detector. The authors also gratefully
acknowledge the support of the research by these agencies and by the
Australian Research Council, the Natural Sciences and Engineering
Research Council of Canada, the Council of Scientific and Industrial
Research of India, the Department of Science and Technology of
India, the Spanish Ministerio de Educacion y Ciencia, The National
Aeronautics and Space Administration, the John Simon Guggenheim
Foundation, the Alexander von Humboldt Foundation, the Leverhulme
Trust, the David and Lucile Packard Foundation, the Research
Corporation, and the Alfred P.\ Sloan Foundation.

\end{document}